\documentclass[conference,a4paper]{IEEEtran}
\IEEEoverridecommandlockouts
\usepackage{stfloats,color}
\usepackage{amsfonts}
\usepackage[cmex10]{amsmath}
\usepackage{graphicx}
\usepackage{setspace}
\usepackage{cite}
\usepackage{array}
\usepackage{algorithmic}
\usepackage{mdwmath}
\usepackage{mdwtab}
\usepackage[center]{caption}
\usepackage{fixltx2e}
\usepackage{url}
\usepackage{times}
\usepackage{epsfig}
\usepackage{latexsym}
\usepackage{epstopdf}
\usepackage{verbatim}
\usepackage{units}
\usepackage{amsthm}
\usepackage{mdwlist}

\usepackage{soul}\graphicspath{{./figures/}}
\newtheorem{thm}{Theorem}
\newtheorem*{prof}{Proof}

\newtheorem{lemm}{Lemma}
\begin{document}

\sloppy

\title{The Deterministic Multicast Capacity of 4-Node Relay Networks }

\author{\IEEEauthorblockN{ Ahmed A. Zewail, M. Nafie}
\IEEEauthorblockA{Wireless Intelligent Networks Center (WINC)\\ Nile University \\ Giza, Egypt \\  
Email: ahmed.zewail@nileu.edu.eg, \\
\ mnafie@nileuniversity.edu.eg}
\and
\IEEEauthorblockN{ Y. Mohasseb}
\IEEEauthorblockA{ Department of Communications \\ The Military Technical College \\ Cairo, Egypt 11331\\
Email: mohasseb@ieee.org}
\and
\IEEEauthorblockN{H. El Gamal}
\IEEEauthorblockA{ECE Department\\ The Ohio State University\\ Columbus, OH \\
Email: helgamal@ece.osu.edu}
\thanks{This paper was made possible by NPRP grant $\#$ 4-1119-2-427 from the Qatar National Research Fund (a member of Qatar Foundation). The statement made herein are solely responsibility of the authors.}}




\maketitle
\begin{abstract}
In this paper, we completely characterize the deterministic capacity region of a four-node relay network with no direct links between the nodes, where each node communicates with the three other nodes via a relay. Towards this end, we develop an upper bound on the deterministic capacity region, based on the notion of a one-sided genie. To establish achievability, we use the detour schemes that achieve the upper bound by routing specific bits via indirect paths instead of sending them directly.  
\end{abstract}

\section{Introduction}
Avestimehr, Diggavi, and Tse presented a deterministic channel model which captures several key features of multiuser wireless communication \cite{avestimehr2011wireless}. They consider a model for a wireless network with nodes connected by such deterministic channels, and present an exact characterization of the end-to-end capacity when there is a single source and a single destination and an arbitrary number of relay nodes. The deterministic model simplifies the wireless network interaction model by eliminating the noise and allows us to focus on the interaction between signals. They obtained sharp results on the deterministic capacity region of single relay channel and diamond channel, and used these results to find an approximate capacity region for each corresponding Gaussian network, where the approximation error can be typically ignored in the high Signal-to-Noise Ratio (SNR) regime. In  \cite{avestimehr2009capacity}, the authors studied the multi-pair bidirectional relay network which is a generalization of the bidirectional relay channel. They examined this problem in the context of the deterministic channel model and characterized its capacity region completely in both full-duplex and half-duplex cases. Also, they showed that the capacity can be achieved by a simple equation-forwarding strategy and illustrated some structures on the signal levels. In \cite{hassibi2009approximate}, the authors proposed a transmission strategy for the Gaussian two-pair two-way full-duplex relay network and found an approximate characterization of the capacity region, based on insights from a recently proposed deterministic channel model.
In  \cite{mokhtar2010deterministic}, the authors developed a new upper bound on the deterministic capacity region, based on the notion of a one-sided genie and completely characterized the multicast deterministic capacity of the two user bidirectional half duplex wireless relay network with only private messages. They also constructed novel detour schemes that achieve the upper bound by routing the bits intended for a certain receiver through the network rather than sending it directly. Then, the authors of \cite{chaaban2011capacity} studied the linear shift deterministic Y-channel, and derived an upper bound which, when combined with the cut-set bounds, provide an outer bound on the capacity region.\\
This paper extends this line of work by completely characterizing the deterministic capacity region of a four-node relay network, where each node has the ability to exchange private messages with the three other nodes. We believe this is an important step towards characterizing the K-node relay network. We consider a multi-hop scenario with no direct links between the communicating nodes. We focus on the symmetric case where the uplink and downlink channels are reciprocal. One can argue that this symmetry assumption corresponds to a Time Division Duplex (TDD) network with relatively slow fading dynamics. To establish achievability, we construct two different detour schemes where specific bits are routed through indirect paths towards their destinations. The choice of the detour scheme depends on the specific rates of the network. However, we show constructively that at least one detour exist which achieves the upper bound. 
It should be noted that these detour schemes are not identical to the one developed in \cite{mokhtar2010deterministic} due to the different nature of the network. We then show that our achievability schemes meet the upper bound of the capacity region via using the notion of a one-sided genie.\newline
The rest of the paper is organized as follows. In Section II, we describe our main assumptions and state the main result of our work. Our new achievability schemes are detailed in Section III. In Section IV, we show the development of the upper bound based on the notation of single sided genie. Section V reports numerical examples to illustrate our achievability schemes. Finally, Section VI presents our conclusions.
\section{Main Result}
Our work focuses on the network shown in Fig.1. It consists of 4 nodes which communicate only through the relay, with no direct link between them. Furthermore, each node can exchange private messages with the three other nodes. We assume that the channel between each node and the relay is reciprocal. In a deterministic channel model \cite{avestimehr2011wireless}, this means that the channel gains, expressed as the number of levels between node $i$ and the relay, is the same in the uplink and downlink i.e. ($ n_{iR} = n_{Ri} = n_i$) where $n_i =\lceil 0.5 \log SNR \rceil$. We can assume without loss of generality that
$ n_1 \geq n_2 \geq n_3 \geq n_4$, otherwise we can re-label the nodes. \newline
This paper derives the deterministic capacity of this network which is given by the following theorem:
\begin{figure}
\includegraphics[width=0.5\textwidth,height=0.12\textheight]{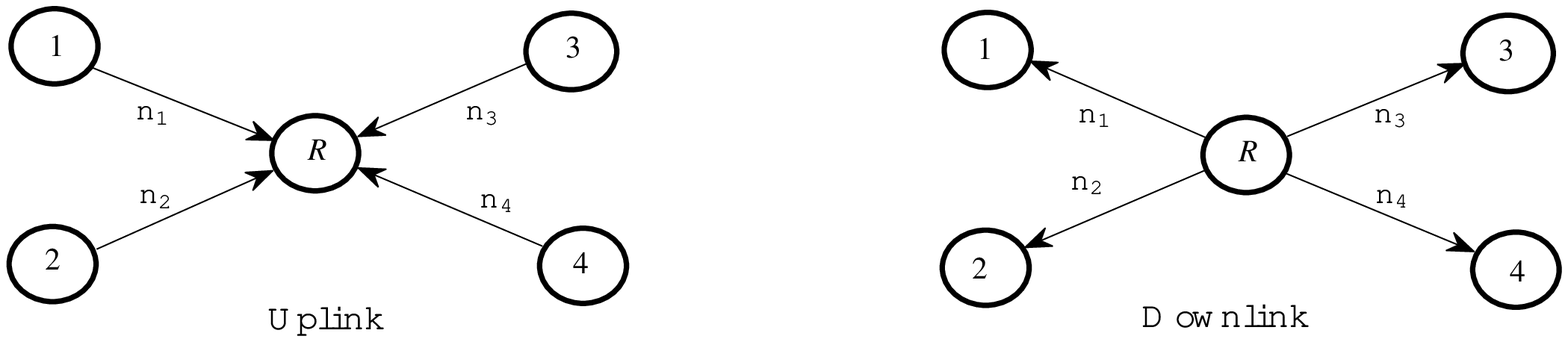}
\centering
\caption{System model}\label{fig:sm}
\end{figure}
\vspace{-0.1 in} 
\begin{thm}\label{upperbound}
The deterministic capacity region of the four-node relay network is given by the rates that satisfy the inequalities in (\ref{DL}), (\ref{UL}) and (\ref{comman}).
\vspace{-0.1 in}
\begin{equation}\nonumber
R_{14}+R_{24}+R_{34}\leq n_4
\end{equation}
\begin{equation}\nonumber
R_{13}+R_{23}+R_{14}+R_{24}+\max(R_{34},R_{43})\leq n_3
\end{equation}
\begin{equation}\nonumber
R_{12}+R_{13}+R_{14}+R_{32}+R_{42}+\max(R_{34},R_{43})\leq n_2
\end{equation}
\begin{equation}\nonumber
R_{12}+R_{13}+R_{14}+R_{23}+R_{43}+\max(R_{24},R_{42})\leq n_2
\end{equation}
\begin{equation}\label{DL}
R_{12}+R_{13}+R_{14}+R_{24}+R_{34}+\max(R_{23},R_{32})\leq n_2
\end{equation}
\begin{equation}\nonumber
R_{41}+R_{42}+R_{43}\leq n_4
\end{equation}
\begin{equation}\nonumber
R_{31}+R_{32}+R_{41}+R_{42}+\max(R_{34},R_{43})\leq n_3
\end{equation}
\begin{equation}\nonumber
R_{21}+R_{31}+R_{41}+R_{23}+R_{24}+\max(R_{34},R_{43})\leq n_2
\end{equation}
\begin{equation}\nonumber
R_{21}+R_{31}+R_{41}+R_{32}+R_{34}+\max(R_{24},R_{42})\leq n_2
\end{equation}
\begin{equation}\label{UL}
R_{21}+R_{31}+R_{41}+R_{42}+R_{43}+\max(R_{23},R_{32})\leq n_2
\end{equation}
\vspace{-.3 in}
\begin{multline}\label{comman}
\max\{(R_{12}+R_{13}+R_{42}+R_{43}),(R_{21}+R_{31}+R_{24}+R_{34})\}\\+\max(R_{23},R_{32})+\max(R_{14},R_{41})\leq n_1
\\
\max\{(R_{12}+R_{14}+R_{32}+R_{34}),(R_{21}+R_{41}+R_{23}+R_{43})\}\\+\max(R_{13},R_{31})+\max(R_{24},R_{42})\leq n_1
\\
\max\{(R_{13}+R_{14}+R_{23}+R_{24}),(R_{31}+R_{41}+R_{32}+R_{42})\}\\+\max(R_{12},R_{21})+\max(R_{34},R_{43})\leq n_1
\end{multline}
where $R_{ij}$ is the rate from node $i$ to node $j$ .
\end{thm}
\vspace{-0.1 in} 
\section{Achievability}\label{Scheme}
\vspace{-0.07 in} 
In this section, we prove the achievability of all rate tuples satisfying Theorem \ref{upperbound} using one of two schemes: either the Simple Ordering Scheme (SOS) or a Detour Scheme (DS) which attempts to find an equivalent network with modified rates that can then be accommodated by the SOS. 
\subsection {The Simple Ordering Scheme (SOS)}
The main idea behind the SOS is that if two nodes $i$ and $j$ wish to exchange a single bit, then they will need only use one channel level. In the uplink phase, each node will send its bit over the assigned channel level, therefore the relay will receive $x_{ij} \oplus x_{ji}$. In the downlink phase, where the relay does not need to decode each bit individually, it can simply broadcast $x_{ij} \oplus x_{ji}$, since node $i$ knows $x_{ij}$, it can decode $x_{ji}$, and vice verse for node $j$.  
\subsubsection {SOS for the Downlink}
The messages to be transmitted are divided into four segments. The first segment contains messages intended for node 4 and it will be constructed as follows. Let $\zeta_k = \min(R_{4k},R_{k4})$  and $\eta_k = \max(R_{4k},R_{k4})$ for $k \in \{1,2,3\}$.
We XOR the first $\zeta_k$ bits in $R_{4k}$ with the corresponding bits in $R_{k4}$. This results in a segment of size ($\zeta_1 + \zeta_2 + \zeta_3$). After inserting the XORed bits as previously mentioned, we append with any remaining bits to be transmitted to 4, i.e. ($\eta_k - \zeta_k$) bits from $R_{k4}$ if $R_{k4} > R_{4k}$, otherwise all bits intended for node 4 will have already been included in the segment. \newline
The second, third, and fourth segments are dedicated to messages intended for nodes 3, 2, and 1, respectively and are constructed in the same manner. In the higher segments, only the remaining bits in each stream that were not included in previous segments are considered. 
\subsubsection {SOS for the Uplink}
Since the relay is not required to decode all the received bits individually in the uplink phase. It is sufficient to provide the relay with the XORed bits resulting from signal level interactions. These XORed bits will be used 'as-is' in downlink phase. The relay needs only to re-order these bits to match the downlink segments described earlier.\\
In principle, this scheme is the same as the ones used in \cite{avestimehr2009capacity} and \cite{mokhtar2010deterministic}.
\begin{lemm}\label{lemm:simple}
The Simple Ordering Scheme (SOS) achieves all the integral rate tuples in the intersection between the capacity regions stated in Theorem \ref{upperbound} and the following extra conditions:
\small
\begin{equation}\label{eqndet321}
\max(R_{23}+R_{34}+R_{42},R_{32}+R_{24}+R_{43})+R_{12}+R_{13}+R_{14}\leq n_2
\end{equation}
\begin{equation}\label{eqndet322}
\max(R_{23}+R_{34}+R_{42},R_{32}+R_{24}+R_{43})+R_{21}+R_{31}+R_{41}\leq n_2
\end{equation}
\begin{equation}\label{eqndet31}
R_{ij}+R_{jk}+R_{ki}+\max(R_{li}+R_{lj}+R_{lk},R_{il}+R_{jl}+R_{kl})\leq n_1
\end{equation}
\begin{equation}\label{eqndet41}
R_{ij}+R_{jk}+R_{kl}+R_{li}+\max(R_{jl},R_{lj})+\max(R_{ik},R_{ki})\leq n_1
\end{equation}
\normalsize
for any $\{i,j,k,l\}$ $\in$   $\{1,2,3,4\}$.
\end{lemm}
\hspace{-.2 in }
\textbf{Proof Sketch:} The SOS can only work if node $i$ is able to transmit or receive all its data on the available number of levels $n_i$, which means that each segment can be accommodated in the corresponding channel. The proof depends on finding the size of each of the four segments, and applying this condition to it in both uplink and downlink phases. For example, the size of segment 4 ($SS_4$) in uplink is given by:
\begin{equation}\nonumber
SS_4= R_{41}+R_{42}+R_{43}\leq n_4
\end{equation}  
Proceeding towards segment 3, we calculate the size ($SS_3$), then apply the condition
\begin{equation}\nonumber
SS_3+SS_4=R_{31}+R_{32}+R_{41}+R_{42}+\max(R_{34},R_{43})\leq n_3
\end{equation}   
Finally, the set of all inequalities obtained from the integer rate tuple constitute achievability region for the SOS, by comparing these inequalities to the ones stated in Theorem \ref{upperbound}, we find the set of extra conditions stated in (\ref{eqndet321})-(\ref{eqndet41}).\newline    
An important note here is that the extra conditions in equations (\ref{eqndet321})-(\ref{eqndet41}), all contain "cycles" in the network as explained in the following subsection. We believe this is an important observation that can later allow determining the capacity region of the K-node network. 
\vspace{-0.05 in}   
\subsection {The Detour Scheme (DS)}
\vspace{-0.05 in} 
If an achievable rate tuple violates any of the extra conditions stated in (\ref{eqndet321})-(\ref{eqndet41}), the SOS will not achieve this tuple. In such cases a detour scheme will be used to convert the network into an equivalent one, where (\ref{eqndet321})-(\ref{eqndet41}) are satisfied, to which we can apply the SOS.\newline
Before explaining the details of our detour scheme, we first observe that the set of extra conditions represented by (\ref{eqndet321})-(\ref{eqndet31}) contain a 3-node cycle corresponds to the data flow in the first three term of LHS. In contrast, conditions represented by (\ref{eqndet41}) contain two 3-node cycles. For example if $\max (R_{jl},R_{lj})=R_{jl}$ and  $\max (R_{ik},R_{ki})=R_{ki}$, then the 3-node cycles are $i, j, k$ obtained from rates  $R_{ij}, R_{jk}$ and $R_{ki}$, and the cycle  $i, j, l$ obtained from rates  $R_{ij}, R_{jl}$ and $R_{li}$. The notion of these cycles will be important in defining our detour scheme. 
\newline
Note that an achievable rate tuple may violate more than one of the conditions expressed by (\ref{eqndet321})-(\ref{eqndet41}). We define the \textbf{Maximum Gap Condition (MGC)} as the condition having the maximum difference between the RHS and LHS of the inequalities over all conditions expressed by (\ref{eqndet321})-(\ref{eqndet41}).\\  
Now we have two detour schemes, depending on the MGC:
\subsubsection* {Detour Scheme 1 (DS 1)}
The MGC is in the form of (\ref{eqndet321}), (\ref{eqndet322}) or (\ref{eqndet31}) 
for a certain $\{i,j,k,l\}$. In this case, the detour will be performed through the 3-node cycle represented by the MGC. To simplify the notation, we assume without loss of generality ($ R_{il}+R_{jl}+R_{kl}) \leq (R_{li}+R_{lj}+R_{lk})$, hence the MGC in the form of (\ref{eqndet31}) can be written as:
\begin{equation}\nonumber
R_{ij}+R_{jk}+R_{ki}+R_{li}+R_{lj}+R_{lk}> n_1
\end{equation}
Now we need to reduce the rates of the left hand side by subtracting $\lambda$, such that the modified rates satisfy:
\begin{equation}\nonumber
(R_{ij}+R_{jk}+R_{ki})-\lambda+R_{li}+R_{lj}+R_{lk}\leq n_1
\end{equation}
The subtracted bits will be transmitted to their respective destinations via alternative paths (detours). Thus all rates along this detour must be increased, while at the same time satisfying the other conditions in Theorem \ref{upperbound} and (\ref{eqndet321})-(\ref{eqndet41}). Considering the reverse cycle, the rates should be modified as: 
\begin{equation}\nonumber
R_{ji}+R_{ik}+R_{kj}\rightarrow R_{ji}+R_{ik}+R_{kj}+2\lambda
\end{equation}
\subsubsection* {Detour Scheme 2 (DS 2)}
The MGC is in the form of (\ref{eqndet41}) for a certain $\{i,j,k,l\}$. In this case, the detour will be performed through the two 3-node cycles represented in the MGC. Again, we assume without loss of generality $\max(R_{jl},R_{lj})+\max(R_{ik},R_{ki})=R_{ik}+R_{lj}$, hence the MGC in the form of (\ref{eqndet41}) can be written as:
\begin{equation}\nonumber
R_{ij}+R_{jk}+R_{kl}+R_{li}+R_{ik}+R_{lj} > n_1
\end{equation}
First, we should define the 3-nodes cycles in the MGC which are
\begin{equation}\nonumber
\begin{aligned}
R_{kl} \rightarrow R_{lj} \rightarrow R_{jk}
& &
R_{kl} \rightarrow R_{li} \rightarrow R_{ik}
\end{aligned}
\end{equation}
We need to reduce LHS, therefore we have to subtract an integer $\alpha$ from LHS such that the reduced rates satisfy:
\begin{equation}\nonumber
R_{ij}+(R_{kl}+R_{li}+R_{ik}+R_{lj}+R_{jk})-\alpha\leq n_1
\end{equation}
The omitted $\alpha$-bits will be transmitted to their respective destinations via alternative paths (detours). Thus all rates along this detour must be increased, while at the same time satisfying the other conditions in Theorem \ref{upperbound} and  (\ref{eqndet321})-(\ref{eqndet41}). Considering the reverse cycles, the rates should be modified as: 
\begin{equation}\nonumber
R_{lk}+R_{ki}+R_{il}+R_{kj}+R_{jl}\rightarrow R_{lk}+R_{ki}+R_{il}+R_{kj}+R_{jl}+2\alpha
\end{equation}
\begin{lemm}
For all integer rate tuples for the 4-node relay network satisfying Theorem \ref{upperbound} and where any of the conditions in Lemma 1 is violated, it is possible to modify the rates using one of the two detour schemes, to find an equivalent network, that can achieve the original rate tuple via alternative paths.
\end{lemm}
\begin{prof}
See the Appendix.  
\end{prof}
\vspace{-.10 in}
\section{Upper bound Based on Single sided genie}
\vspace{-.05 in}
The relay channel can be represented as the combination of Multiple Access channel i.e. (Uplink) and Broadcast channel i.e. (Downlink). In the traditional cut set bounds as shown in \cite{cover2006elements}, we divide the nodes of a network into two sets $S$ and $S^c$ which represent the transmitting and receiving nodes respectively. As was mentioned in \cite{mokhtar2010deterministic}, if all nodes in $S^c$ fully cooperate and share all their side information, we refer to this cooperation as the two sided genie aided bound. As was shown in \cite{mokhtar2010deterministic}, applying this traditional cut set bound to the relay network produces loose bounds, therefore a tighter single sided genie aided upper bound was developed in \cite{mokhtar2010deterministic}.
\subsection{The Downlink Upper Bound}
If we consider the cut on the downlink phase with 
$S=\{i,j\}$ and $S^c=\{k,l\}$, the two sided genie, cut bound will be:
\begin{equation}\nonumber
R_{ki}+R_{li}+R_{kj}+R_{lj}\leq \max(n_i,n_j)
\end{equation}    
However, with the one sided genie presented in \cite{mokhtar2010deterministic}, we assume that the genie transfers only all data of node $i$ to node $j$ i.e.($R_{ij}$), therefore the data sent from node $j$ to node $i$ i.e.($R_{ji}$) is not known at node $i$ a-priori. This results in a tighter inequality: 
\begin{equation}\nonumber
R_{ki}+R_{li}+R_{kj}+R_{lj}+R_{ji}\leq \max(n_i,n_j)
\end{equation} 
Conversely, if the genie transfers only all the data of node $j$ to node $i$, the bound will be as follows. 
\begin{equation}\nonumber
R_{ki}+R_{li}+R_{kj}+R_{lj}+R_{ij}\leq \max(n_i,n_j)
\end{equation} 
These two conditions can be combined as follows.
\begin{equation}\nonumber
R_{ki}+R_{li}+R_{kj}+R_{lj}+\max(R_{ij},R_{ji})\leq \max(n_i,n_j)
\end{equation}
For $S=\{i,j,k\}$ and $S^c=\{l\}$, if we assume that the genie transfers all data of node $i$ to nodes $j$ and $k$ i.e. $R_{ij}$ and $R_{ik}$ , and all data of node $j$ to node $k$ i.e. $R_{jk}$ , therefore the data sent from node $k$ to nodes $i$ and $j$ i.e. $R_{ki}$ and $R_{kj}$ is not known at nodes $i$ and $j$, and the data sent from node $j$ to node $i$ i.e. $R_{ji}$ is not known at node $i$.
This results in a tighter inequality as follows:
\begin{equation}\nonumber
R_{li}+R_{lj}+R_{lk}+R_{ji}+R_{ki}+R_{kj}\leq \max(n_i,n_j,n_k)
\end{equation} 
The previous inequality represents only one of the different genie orders, which must be taken into account to characterize the upper bound.\newline 
It should be mentioned that for $S=\{i\}$ i.e. (cut contains only one node), the single sided genie bound coincides with the traditional two sided genie. In contrast, we need the cut around the relay, and the two sided genie is not useful in this case, where the one sided genie bound depends only on the different genie orders. For example, if assume the genie order $i \rightarrow j \rightarrow k \rightarrow l$, this means the genie transfers all data of node $i$ to nodes $j, k$ and $l$, then the data of node $j$ to nodes $k$ and $l$, and finally, it transfers the data from node $k$ to node $l$, there for the one sided genie bound in this case will be:
\begin{equation}\nonumber
R_{li}+R_{lj}+R_{lk}+R_{ki}+R_{kj}+R_{ji}\leq n_1
\end{equation} 
To get the upper bound, we should take all possible cuts and genie orders. Finally, we can write the downlink bounds as stated in (\ref{DL}) and (\ref{comman}). 
\vspace{-0.05 in}
\subsection{The Uplink Upper bound}
\vspace{-0.05 in}
For the uplink, if we take $S=\{i,j\}$ and $S^c=\{k,l\}$. If we assume that the genie transfers only all the data of node $i$ to node $j$ i.e.($R_{ij}$), this help the relay decode the data transmitted to the nodes of $S^c$, therefore the bounds will be as follows.
\begin{equation}\nonumber
R_{ik}+R_{il}+R_{jk}+R_{jl}+R_{ji}\leq \max(n_i,n_j)
\end{equation} 
Again, if the genie order is exchanged, then the bound will be: 
\begin{equation}\nonumber
R_{ik}+R_{il}+R_{jk}+R_{jl}+R_{ij}\leq \max(n_i,n_j)
\end{equation} 
These two conditions can be combined as follows.
\begin{equation}\nonumber
R_{ik}+R_{il}+R_{jk}+R_{jl}+\max(R_{ij},R_{ji})\leq \max(n_i,n_j)
\end{equation} 
We proceed in a similar manner with all possible cuts and all genie orders. Finally, we can write the uplink bounds as stated in (\ref{UL}) and (\ref{comman}). It should be noted that (\ref{comman}) is common between the uplink and the downlink bounds, this results from the cut around the relay, where the bounds depend only on the genie order, therefore from genie order $i \rightarrow j \rightarrow k \rightarrow l$ in uplink phase, we get the same bound from the reversed genie order in downlink phase i.e.($l \rightarrow k \rightarrow j \rightarrow i$). Also, it should be mentioned that some cuts does not give a new constraint on the capacity region, therefore we did not state them in Theorem 1. For example, the bound we get from the cut $S=\{3\}$ is already satisfied from the bound that we get from the cut $S=\{3,4\}$.    
\section{Numerical examples}
Consider a reciprocal network with channel gains $(n_1, n_2, n_3, n_4) = (7, 6, 5, 4)$ and the rate tuple $R= \{R_{12},R_{13},R_{14},R_{21},R_{23},R_{24},R_{31},R_{32},R_{34},R_{41},R_{42},R_{43}\}$.
\subsubsection*{Example 1}
$R=\{2,0,0,0,0,2,1,1,1,1,0,0\}$ which satisfies Theorem \ref{upperbound}, but violates conditions from Lemma 1 and the MGC is: $R_{12}+R_{24}+R_{41}+R_{31}+R_{32}+R_{34}=7+1$. 
This example will solved by \textbf{DS 1} as follows. $R_{24}\rightarrow R_{24}-1, R_{21}\rightarrow R_{21}+1$ and $R_{14}\rightarrow R_{14}+1$. Now we have an equivalent network with the new rate tuple $\grave{R}=\{2,0,1,1,0,1,1,1,1,1,0,0\}$, that satisfies Theorem \ref{upperbound} and (\ref{eqndet321})-(\ref{eqndet41}). 
\subsubsection*{Example 2}
$R=\{0,0,2,1,0,1,1,2,0,0,0,2\}$ which satisfies Theorem \ref{upperbound}, but violates conditions from Lemma 1 and the MGC is: $R_{14}+R_{43}+R_{32}+R_{21}+R_{31}+R_{24}=7+2$.\newline 
This example will solved by \textbf{DS 2} as follows. $R_{43}\rightarrow R_{43}-2, R_{41}\rightarrow R_{41}+1, R_{42}\rightarrow R_{42}+1, R_{13}\rightarrow R_{13}+1$ and $R_{23}\rightarrow R_{23}+1$ as shown in Fig.2. Now we have an equivalent network with the new rate tuple $\grave{R}=\{0,1,2,1,1,1,1,2,0,1,1,0\}$, that satisfies Theorem \ref{upperbound} and (\ref{eqndet321})-(\ref{eqndet41}).
\begin{figure}
\includegraphics[width=0.45\textwidth,height=0.17\textheight]{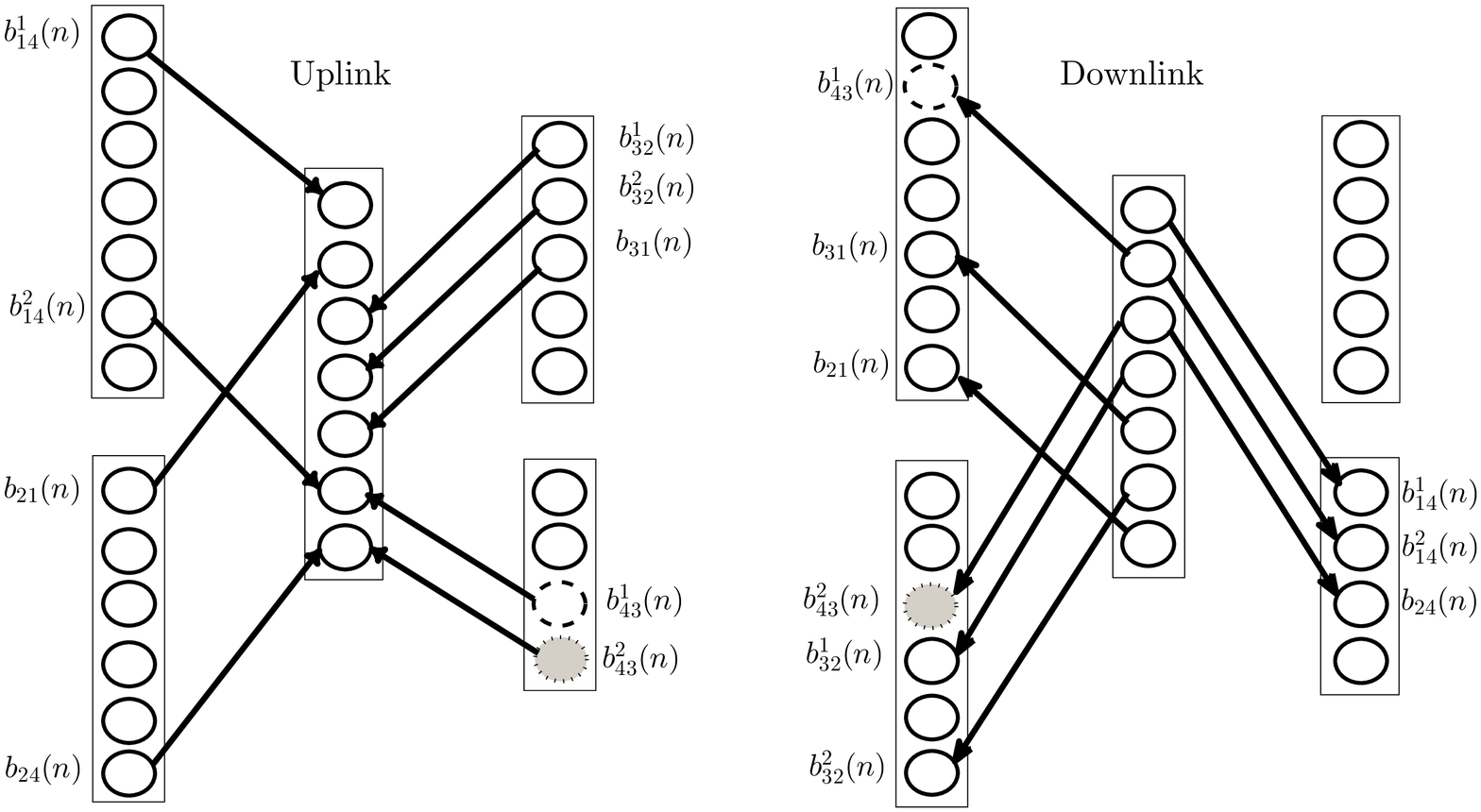}
\centering
\includegraphics[width=0.45\textwidth,height=0.17\textheight]{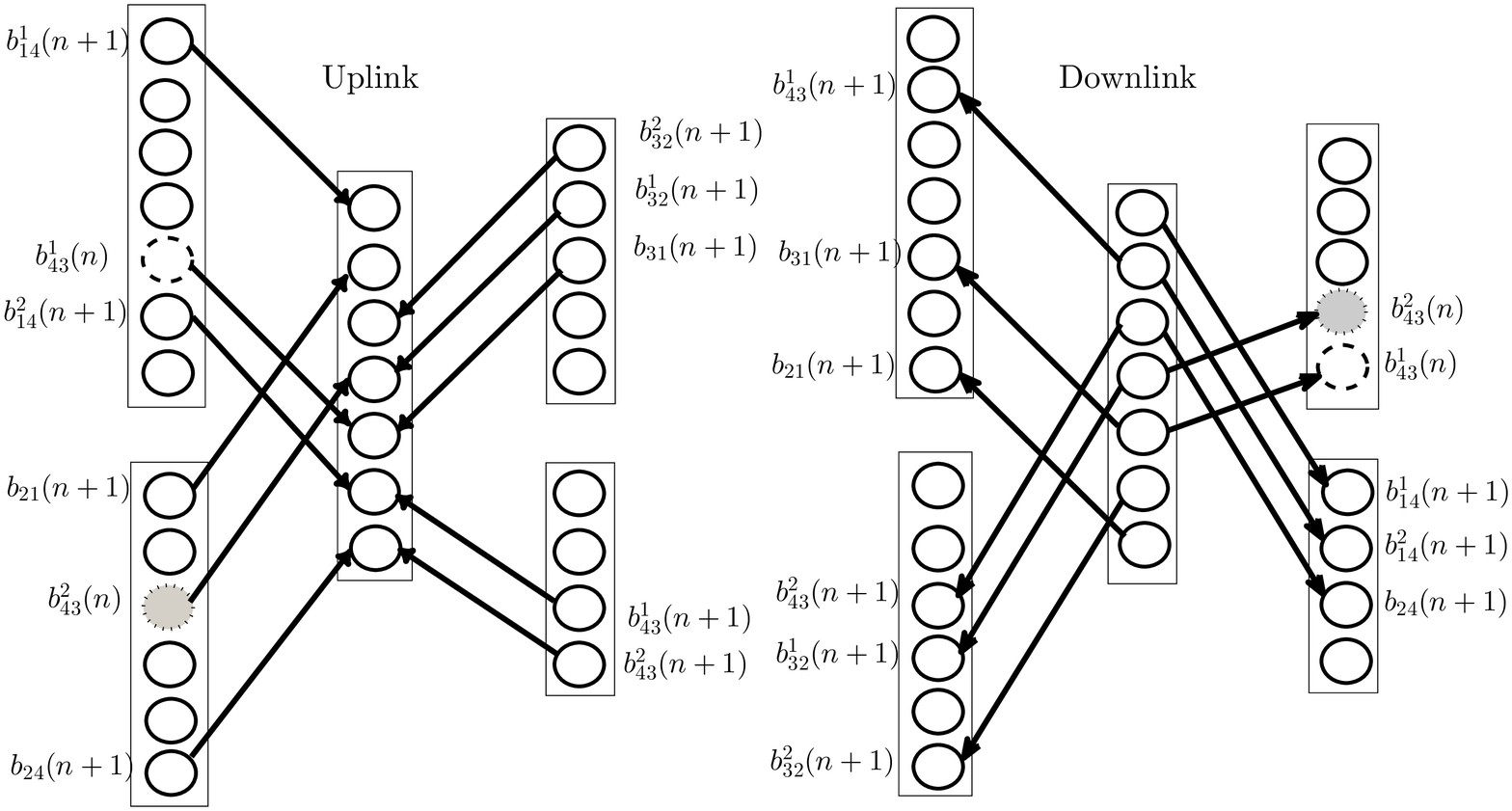}
\centering
\caption{Example on the Detour Scheme 2}
\end{figure}
\section{Conclusions}
In this work, we characterized of the deterministic capacity region of a four-node relay network with no direct links. Our work results in defining a tighter upper bound based on the notion of single sided genie, and provides two schemes of coding across levels and time to achieve this bound. Our current investigations target the generalization to the K-node network with common information, and using the insights gleaned from this deterministic channel model to obtain coding schemes and approximate capacity results for the corresponding Gaussian network. We argue that we can drive the upper bound of the deterministic capacity region in a systematic way based on the notion of the single sided genie similar to that we followed to in this paper, also the extra conditions to apply SOS would contain multiples of 3-node cycles. If any of these extra conditions is violated, we can use detour schemes to convert the network into an equivalent one that satisfies the upper bound and the extra conditions, then we can apply SOS to it. We also argue that all detours would be applied via one or more 3-node cycles. We had tested our arguments on 5-node relay network, and the results indicate that they are correct. 
\vspace{-0.07 in} 
\section*{Appendix}
\vspace{-.07 in}
\subsubsection*{DS 1}
Assume without loss of generality, a condition in the form of (\ref{eqndet31}) where $\{i,j,k,l\}=\{1,2,3,4\}$ is the MGC violated by $\lambda$ bits, then we can write:
\begin{equation}\nonumber
R_{13}+R_{34}+R_{41}+R_{12}+R_{32}+R_{42}=n_1+\lambda
\end{equation}
From the upper bound (\ref{comman}), we have
\begin{equation}\nonumber
R_{31}+R_{34}+R_{32}+R_{41}+R_{42}+R_{12}\leq n_1
\end{equation}
\begin{equation}\nonumber
R_{41}+R_{42}+R_{43}+R_{12}+R_{13}+R_{32}\leq n_1
\end{equation}
\begin{equation}\nonumber
R_{13}+R_{12}+R_{14}+R_{32}+R_{34}+R_{42}\leq n_1
\end{equation}
By comparing the above conditions with the MGC, we get 
\begin{equation}\label{eqn: R1icondUB}
\begin{aligned}
R_{13} \geq R_{31}+\lambda
& & R_{34} \geq R_{43}+\lambda
& & R_{41} \geq R_{14}+\lambda
\end{aligned}
\end{equation} 
From the other extra SOS conditions, we have
\begin{equation}\nonumber
R_{21}+R_{13}+R_{34}+R_{41}+R_{32}+R_{42}\leq n_1+\lambda
\end{equation}
\begin{equation}\nonumber
R_{13}+R_{23}+R_{34}+R_{41}+R_{12}+R_{42}\leq n_1+\lambda
\end{equation}
\begin{equation}\nonumber
R_{13}+R_{34}+R_{24}+R_{41}+R_{12}+R_{32}\leq n_1+\lambda
\end{equation}
By comparing the above conditions with the MGC, we get
\begin{equation}\label{eqn: R1icondSOS}
\begin{aligned}
R_{12} \geq R_{21}
& & R_{32} \geq R_{23}
& & R_{42} \geq R_{24}
\end{aligned}
\end{equation}
Now, we will apply the DS 1, we detour $\lambda$ bits from node 4 to node 1 via node 3, therefore the new rates are as follows.
\begin{equation}\nonumber
\begin{aligned}
R_{41} \rightarrow R_{41}-\lambda
& & R_{43} \rightarrow R_{43}+\lambda
& & R_{31} \rightarrow R_{31}+\lambda
\end{aligned}
\end{equation}
Then, we need to check the upper bound and the SOS extra conditions for this new rate tuple. It can be shown from (\ref{eqn: R1icondUB}) and (\ref{eqn: R1icondSOS}) that the conditions stated in Theorem \ref{upperbound} and Lemma 1 are now satisfied.\\
It should be mentioned that if the MGC is in the form of (\ref{eqndet321}) or (\ref{eqndet322}), the proof will follow the same steps but we would compare the MGC with the conditions restricted by $n_2$ in Theorem \ref{upperbound}.  
\subsubsection*{DS 2}
Assume without loss of generality, a condition in the form of (\ref{eqndet41}) where $\{i,j,k,l\}=\{1,2,3,4\}$ is the MGC violated by $\lambda$ bits, then we can write:
\begin{equation}\nonumber
R_{12}+R_{23}+R_{34}+R_{41}+R_{42}+R_{13}=n_1+\lambda
\end{equation} 
This MGC contains two 3-node cycles as follows.
\begin{equation}\nonumber
\begin{aligned}
R_{34}\rightarrow R_{42}\rightarrow R_{23}
& & &
R_{34}\rightarrow R_{41}\rightarrow R_{13}
\end{aligned}
\end{equation}
From the upper bound (\ref{comman}), we can get
\begin{equation}\nonumber
R_{43}+R_{41}+R_{42}+R_{12}+R_{13}+R_{23}\leq n_1
\end{equation} 
By comparing this condition with the MGC, we can get
\begin{equation}\label{eqn: R4condUB}
R_{34} \geq R_{43}+\lambda
\end{equation}
From the extra SOS conditions, we have
\begin{equation}\label{eq1}
R_{34}+R_{42}+R_{23}+R_{12}+R_{13}+R_{14}=n_1+\beta
\end{equation}
By subtracting this condition from the MGC, we can get:
\begin{equation}\label{eqn: R4condSOS1}
R_{41}=R_{14}+\lambda-\beta
\end{equation}
Also, from the upper bound (\ref{comman}), we can get
\begin{equation}\nonumber
R_{12}+R_{13}+R_{14}+R_{23}+R_{24}+R_{34}\leq n_1
\end{equation}
and by comparing with (\ref{eq1}), we get 
\begin{equation}\label{eqn: R4condSOS2}
R_{42}\geq R_{24}+\beta
\end{equation}   
Again, from the extra SOS conditions, we have
\begin{equation}\label{eq2}
R_{34}+R_{41}+R_{13}+R_{12}+R_{42}+R_{32}=n_1+\gamma
\end{equation}
By subtracting this condition from the MGC, we can get
\begin{equation}\label{eqn: R4condSOS3}
R_{23}=R_{32}+\lambda-\gamma
\end{equation}
Also, from the upper bound (\ref{comman}), we can get
\begin{equation}\nonumber
R_{31}+R_{32}+R_{34}+R_{41}+R_{42}+R_{12}\leq n_1
\end{equation}
and by comparing with (\ref{eq2}), we get  
\begin{equation}\label{eqn: R4condSOS4}
 R_{13}\geq R_{31}+\gamma
\end{equation}
where $\lambda=\beta+\gamma$.\\
Now, we will apply DS 2, we detour $\lambda$ bits from node 3 to node 4 via nodes 2 and 1. Therefore, the new rates become: 
\begin{equation}\nonumber
\begin{aligned}
& & &  R_{34} \rightarrow R_{34}-\lambda \\ 
& &
R_{32} \rightarrow R_{32}+\beta
& &
R_{24} \rightarrow R_{24}+\beta
\\
& &
R_{31} \rightarrow R_{31}+\gamma
& &
R_{14} \rightarrow R_{14}+\gamma
\end{aligned}
\end{equation}
Then, we need to check the upper bound and the SOS extra conditions for the new rate tuple. It can be shown from (\ref{eqn: R4condUB}), (\ref{eqn: R4condSOS1}), (\ref{eqn: R4condSOS2}), (\ref{eqn: R4condSOS3}) and (\ref{eqn: R4condSOS4}) that conditions stated in Theorem \ref{upperbound} and Lemma 1 are now satisfied. 
\vspace{-.1 in} 
\bibliographystyle{IEEEtran}
\bibliography{IEEEabrv,DiversityLib}
\vspace{-.15 in} 
\end{document}